\documentclass{appolb-blank}  

\usepackage{graphicx}

\usepackage{latexsym,amssymb,amsmath,amsfonts,epsfig,mathtools}  

\newcommand{\beq}{\begin{equation}}   
\newcommand{\eeq}{\end{equation}}
\newcommand{\beqa}{\begin{eqnarray}}
\newcommand{\eeqa}{\end{eqnarray}}
\newcommand{\beqNO}{\begin{equation*}}
\newcommand{\eeqNO}{\end{equation*}}
\newcommand{\beqaNO}{\begin{eqnarray*}}
\newcommand{\eeqaNO}{\end{eqnarray*}}
\newcommand{\bsubeqs}{\begin{subequations}}
\newcommand{\esubeqs}{\end{subequations}}
\newcommand\fraclarge[2]{\displaystyle{\frac{\textstyle #1}{\textstyle #2}}}

\begin{document}

\eqsec  

\title{Solutions of the bosonic master-field equation\\
       from a supersymmetric matrix model
 \\}   

\author{F.R. Klinkhamer
\address{
Institute for Theoretical Physics,
Karlsruhe Institute of Technology (KIT),\\
76128 Karlsruhe, Germany\\
\texttt{frans.klinkhamer@kit.edu}
}\\ 
}
\maketitle
\begin{abstract}  
It has been argued that
the bosonic large-$N$ master field of the IIB matrix model
can give rise to an emergent classical spacetime.
In a recent paper, we have obtained solutions of
a \emph{simplified} bosonic master-field equation from a related
matrix model. In this simplified equation,
the effects of dynamic fermions were removed.
We now consider the \emph{full} bosonic master-field equation
from a related supersymmetric matrix model for
dimensionality $D=3$ and matrix size $N=3$.
In this last equation,
the effects of dynamic fermions are included.
With an explicit realization of the random constants
entering this algebraic equation, we establish the existence of
nontrivial solutions.
The small matrix size, however, does not allow us to
make a definitive statement as to the appearance of a
diagonal/band-diagonal structure in the obtained matrices.
\end{abstract}

\vspace{-125mm}
\noindent Acta Phys. Pol. B 52, 1339 (2021)
\hfill arXiv:2106.07632
\vspace{+125mm}



\section{Introduction}
\label{sec:Intro}

The IIB matrix model~\cite{IKKT-1997,Aoki-etal-review-1999}
has been proposed as a
nonperturbative formulation of type-IIB superstring theory.
Recently, the corresponding large-$N$ bosonic master
field~\cite{Witten1979,GreensiteHalpern1983}
has been suggested as the possible source of an emerging
classical spacetime~\cite{Klinkhamer2020-master}
(see also Ref.~\cite{Klinkhamer2020-reg-bb-IIB-m-m}
for a follow-up paper on cosmology
and Ref.~\cite{Klinkhamer2021-APPB-review}  for a review).

The task, now, is to solve the relevant bosonic master-field equation.
Preliminary results have shown that there may appear a bosonic
master-field solution, whose matrices have an approximate
diagonal/band-diagonal structure~\cite{Klinkhamer2021-first-look}.
These results were, however, obtained from a \emph{simplified}
bosonic master-field equation with dynamical effects of the
fermions removed altogether.

In the present paper, we will consider
the \emph{full} bosonic master-field equation
with dynamical effects of the fermions included.
There are, then, two crucial questions.
First, does the full bosonic master-field equation
still have solutions?
Second, assuming the existence of a solution,
do the fermion effects preserve 
the diagonal/band-diagonal structure
found from the previous simplified equation?

It is well known that the fermions of the IIB matrix model give rise  
to a Pfaffian, which is extremely difficult to
calculate symbolically~\cite{KrauthNicolaiStaudacher1998,%
NishimuraVernizzi2000-JHEP,AustingWheater2001}.
As a first step towards answering the two
questions of the previous paragraph, we consider
a low dimensionality $D=3$ and a small matrix size $N=3$.
Then, we can give a clear affirmative answer to the first question
on the existence of solutions,
but are not yet able to give a definitive answer to the second question
on a possible diagonal/band-diagonal structure
(even though, the $D=N=3$ results are somewhat encouraging).


\section{Supersymmetric matrix model}
\label{sec:Supersymmetric-matrix-model}

\subsection{General case}
\label{subsec:General-case}

Let us briefly review the IIB matrix
model~\cite{IKKT-1997,Aoki-etal-review-1999},
first allowing for a different number of spacetime dimensions
than ten. We essentially take over the conventions and notation of
Ref.~\cite{KrauthNicolaiStaudacher1998},
except that we write $A^{\mu}$ for the bosonic matrices,
with a directional index $\mu$ running over $\{1,\,\ldots\,,D\}$.
These bosonic matrices, as well as the
fermionic matrices, are $N \times N$ traceless Hermitian matrices.
The partition function for $D>2$ and $N\geq 2$ is then defined as
follows~\cite{IKKT-1997,Aoki-etal-review-1999,KrauthNicolaiStaudacher1998}:
\bsubeqs\label{eq:Z-D-N-F-all-defs}
\beqa
\label{eq:Z-D-N-F}
\hspace*{-2.0mm}
Z_{D,\,N}^{F} &=&
\int \prod_{I=1}^{g}\,\prod_{\mu=1}^{D}\,
\frac{d A_{\mu}^{I}}{\sqrt{2\pi}}\;
e^{\displaystyle{-\,S_{\text{bos}}}[A]}\;
\nonumber\\
\hspace*{-2.0mm}&& \times
\left(\int \prod_{I=1}^{g}\,\prod_{\alpha=1}^{\mathcal{N}}
d\Psi_{\alpha}^{I}\;
e^{\displaystyle{-\,S_{\text{ferm}}[A,\,\Psi]}}\,
\right)^{F}\,,
\\[2mm]
\hspace*{-2.0mm}
\label{eq:Sbos}  
S_{\text{bos}}[A] &=&
-\frac{1}{2}\,\text{Tr}\,
\Big(\big[A^{\mu},\,A^{\nu} \big]\,\big[ A^{\mu},\,A^{\nu} \big]\,\Big)\,,
\\[2mm]
\hspace*{-2.0mm}
\label{eq:Sferm}   
S_{\text{ferm}}[A,\,\Psi] &=&
-\text{Tr}\,
\Big(\, \overline{\Psi}_{\alpha}\,\Gamma^{\mu}_{\alpha\beta}\,
\big[ A^{\mu},\,\Psi_{\beta}\big]\Big)\,,
\\[2mm]
\hspace*{-2.0mm}
\label{eq:Amu-coeff-Psialpha-coeff}
A_{\mu}
&=&A_{\mu}^{I}\,T_{I}\,,
\quad
\Psi_{\alpha} = \Psi_{\alpha}^{I}\,T_{I}
\,,
\\[2mm]
\hspace*{-2.0mm}
\label{eq:trace-T-T}
\text{Tr}\, \big( T_{I} \cdot T_{J} \big)
&=& \frac{1}{2}\;\delta_{I J}\,,
\eeqa
\beqa
\label{eq:g}
\hspace*{-2.0mm}
g &\equiv& N^2-1\,,
\\[2mm]
\hspace*{-2.0mm}
\label{eq:mathcalN}
\mathcal{N} &\equiv& 2^{\,[D/2]} \times
\begin{cases}
 1 \,,   &  \text{for odd}\,D \,,
 \\[2mm]
 1/2 \,,   &  \text{for even}\,D \,,
\end{cases}
\\[2mm]
\label{eq:F-0-or-1}
F &\in&  \big\{ 0,\, 1 \big\}\,,
\eeqa
\esubeqs
where repeated Greek indices are summed over
(corresponding to an implicit Euclidean ``metric'') and   
$F$ is an on/off parameter to include ($F=1$)
or exclude ($F=0$) dynamic-fermion effects.
The square bracket 
in the exponent of \eqref{eq:mathcalN}
stands for the Entier/Floor function
and the factor $1/2$ corresponds to the Weyl projection
[for $D=10$, there is also a reality (Majorana) condition on the
fermions].
The commutators entering the action terms \eqref{eq:Sbos}
and \eqref{eq:Sferm} are defined
by $[X,\,Y]\equiv$ $X \cdot Y - Y \cdot X$
for square ma\-tri\-ces $X$ and $Y$ of equal dimension.

The matrix model \eqref{eq:Z-D-N-F-all-defs} with $F=1$
is supersymmetric for dimensionality
\beq
\label{eq:supersymmetric-D3-D4-D6-D10}
D=3,\,4,\,6,\,10\,,
\eeq
where the field transformations have, for example, been given
by Eq.~(2) in Ref.~\cite{KrauthNicolaiStaudacher1998}.
The expansions \eqref{eq:Amu-coeff-Psialpha-coeff},
for real coefficients $A_{\mu}^{I}$ and
Grassmannian coefficients $\Psi_{\alpha}^{I}$,
use the $N \times N$ traceless Hermitian
$SU(N)$ generators $T_{I}$ with normalization \eqref{eq:trace-T-T}.
Remark also that
we have set the model length scale $ \ell$ to unity,
so that the coefficients $A_{\mu}^{I}$ and $\Psi_{\alpha}^{I}$
are dimensionless.

The Gaussian integrals over the Grassmann variables $\Psi_{\alpha}^{I}$
in \eqref{eq:Z-D-N-F} can be performed analytically,
so that the partition function reduces to a purely bosonic integral,  
\bsubeqs\label{eq:Z-D-N-F-with-Pfaffian-and-Seff-D-N-F}
\beqa
\label{eq:Z-D-N-F-with-Pfaffian}
Z_{D,\,N}^{F} &=&
\int \prod_{I=1}^{g}\,\prod_{\mu=1}^{D}\,
\frac{d A_{\mu}^{I}}{\sqrt{2\pi}}\;
\Big( \mathcal{P}_{D,\,N}[A] \Big)^{F}\;
e^{\displaystyle{-\,S_{\text{bos}}[A]}}
\nonumber\\[1mm]
&=&
\int \prod_{I=1}^{g}\,\prod_{\mu=1}^{D}\,
\frac{d A_{\mu}^{I}}{\sqrt{2\pi}}\;
e^{\displaystyle{-\,S_\text{eff,\,D,\,N}^{F}[A]}}\,,
\\[2mm]
\label{eq:Seff-D-N-F}
S_\text{eff,\,D,\,N}^{F}[A]
&=&
S_{\text{bos}}[A]- F\;\log\,\mathcal{P}_{D,\,N}[A]\,.
\eeqa
\esubeqs
The Pfaffian $\mathcal{P}_{D,\,N}[A]$,
which can be absorbed into the effective action $S_\text{eff}[A]$,
is given explicitly
by a sum over permutations~\cite{KrauthNicolaiStaudacher1998}
or by a sum involving the Levi--Civita symbol.
Concretely, the Pfaffian $\mathcal{P}_{D,\,N}[A]$
is a homogenous polynomial in the bosonic coefficients
$A_{\mu}^{I}$,
where the order $K$,
for the special dimensions \eqref{eq:supersymmetric-D3-D4-D6-D10},
is given by the following
expression~\cite{KrauthNicolaiStaudacher1998,NishimuraVernizzi2000-JHEP}:
\beq
\label{eq:Pfaffian-order-polynomial-K}
K =
\big( D-2\big)\,\big( N^2-1\big)\,.
\eeq
An explicit example of the Pfaffian will be given in
Sec.~\ref{subsec:Particular-case}.
Further discussion of the Pfaffian appears in, e.g.,
Refs.~\cite{KrauthNicolaiStaudacher1998,NishimuraVernizzi2000-JHEP,%
AustingWheater2001}.

The partition function of the genuine 
IIB matrix model~\cite{IKKT-1997,Aoki-etal-review-1999}
has the following parameters in \eqref{eq:Z-D-N-F-all-defs}:
\beq
\label{eq:D-N-F-for-IIB-matrix-model}
\big\{ D,\,  N,\, F \big\}  = \big\{ 10,\,  \infty,\, 1 \big\}\,,
\eeq
and there is a second supersymmetry transformation
in addition to the one mentioned below
\eqref{eq:supersymmetric-D3-D4-D6-D10}.
The large-$N$ limit may require further discussion, but, at this
moment, we just consider $N$ to be large and finite
(for exploratory numerical results, see, e.g.,
Refs.~\cite{KimNishimuraTsuchiya2012,NishimuraTsuchiya2019,%
Anagnostopoulos-etal-2020} and references therein).


\subsection{Particular case}
\label{subsec:Particular-case}

Now consider the matrix model \eqref{eq:Z-D-N-F-all-defs}
with the particular parameters
\beq
\label{eq:D3-N3}
\big\{ D,\,  N,\, F \big\}  = \big\{ 3,\,  3,\, 1 \big\}\,,
\eeq
for which the model has a supersymmetry invariance,
as mentioned in the second paragraph of Sec.~\ref{subsec:General-case}.
The eight generators $T_{I}$ are proportional to the
$3 \times 3$ Gell-Mann matrices $\lambda_{I}$ used in elementary particle physics. Remarkably, there is an explicit result
for the Pfaffian~\cite{KrauthNicolaiStaudacher1998}:  
\beqa
\label{eq:Pfaffian-D3-N3}
\hspace*{-6mm}
\mathcal{P}_{3,\,3}[A] &=&
-\frac{3}{4}\,
\text{Tr}\,\Big(
\big[A^{\mu},\,A^{\nu} \big]\,\big\{ A^{\rho},\,A^{\sigma} \big\}\,
\Big)\;
\text{Tr}\,\Big(
\big[A^{\mu},\,A^{\nu} \big]\,\big\{ A^{\rho},\,A^{\sigma} \big\}\,
\Big)
\nonumber\\[1mm]
\hspace*{-6mm}
&& +
\frac{6}{5}\,\text{Tr}\,
\Big(
A^{\mu}\,\big[ A^{\nu},\,A^{\rho} \big]\,\Big)\
\text{Tr}\,\Big(
A^{\mu}\,\big[ \big\{ A^{\nu},\,A^{\sigma} \big\},\,
              \big\{ A^{\rho},\,A^{\sigma} \big\}\big]\,\Big)\,,
\eeqa
which corresponds to a homogenous eighth-order polynomial
in the bosonic coefficients $A_{\mu}^{I}$.
The expression \eqref{eq:Pfaffian-D3-N3} contains,
in addition to commutators, also anticommutators,
defined by $\{X,\,Y\}\equiv$ $X \cdot Y + Y \cdot X$
for square ma\-tri\-ces $X$ and $Y$ of equal dimension.

Using expression \eqref{eq:Pfaffian-D3-N3} for the Pfaffian,
the effective action is given by
\beq
\label{eq:Seff-D3-N3}
S_\text{eff,\,3,\,3}^{F}[A]
=
S_{\text{bos},\,3,\,3}[A]- F\;\log\,\mathcal{P}_{3,\,3}[A]\,,
\eeq
where $F=0$ removes the effects of dynamic fermions
and $F=1$  includes them.
In a previous paper~\cite{Klinkhamer2021-first-look},
we had simply removed the fermion term in the effective
action but, here, we intend to study it carefully.
Incidentally, the integrals
in \eqref{eq:Z-D-N-F-with-Pfaffian-and-Seff-D-N-F},
for parameters \eqref{eq:D3-N3},
may have convergence problems~\cite{AustingWheater2001},
but our focus will be solely on a type of saddle-point
equation obtained from the effective action \eqref{eq:Seff-D3-N3}.


\section{Bosonic master field}
\label{sec:Bosonic-master-field}

\subsection{Bosonic observables and master field}
\label{subsec:Bosonic-observables-and-master-field}

As our main interest is in the possible recovery of an emerging
classical spacetime~\cite{Klinkhamer2020-master,%
Klinkhamer2020-reg-bb-IIB-m-m,Klinkhamer2021-APPB-review}, we primarily
consider the bosonic observable
\beq \label{eq:IIB-matrix-model-w-observable}
w^{\mu_{1} \,\ldots\, \mu_{m}}
\equiv  
\frac{1}{N}\;
\text{Tr}\,\big( A^{\mu_{1}} \cdots\, A^{\mu_{m}}\big)\,,
\eeq
where the $1/N$ prefactor on the right-hand side is
only for convenience.
Now, arbitrary strings of these bosonic observables
have expectation values
\beqa \label{eq:IIB-matrix-model-w-product-vev}
\hspace*{-3mm}
&&
\langle
w^{\mu_{1}\,\ldots\,\mu_{m}}\:w^{\nu_{1}\,\ldots\,\nu_{n}}\, \cdots\,
w^{\omega_{1}\,\ldots\,\omega_{z}}
\rangle_{D,\,N}^{F}
\nonumber\\[1mm]
&&
= \frac{1}{Z_{D,\,N}^{F}}\,\int dA\,
\big(w^{\mu_{1}\,\ldots\,\mu_{m}}\:w^{\nu_{1}\,\ldots\,\nu_{n}}\, \cdots\,
w^{\omega_{1}\,\ldots\,\omega_{z}}\big)\,
e^{\displaystyle{-\,S_\text{eff,\,D,\,N}^{F}}}\,,
\eeqa
where ``$dA$'' is a short-hand notation of the measure appearing
in \eqref{eq:Z-D-N-F} and $Z_{D,\,N}^{F}$ is defined by the
integral \eqref{eq:Z-D-N-F-with-Pfaffian-and-Seff-D-N-F}.

These expectation values, at large values of $N$,
have a remarkable factorization property:  
\beqa \label{eq:IIB-matrix-model-w-product-vev-factorized}
\hspace*{-3mm}
\langle
w^{\mu_{1}\,\ldots\,\mu_{m}}\:w^{\nu_{1}\,\ldots\,\nu_{n}}\, \cdots\,
w^{\omega_{1}\,\ldots\,\omega_{z}} \rangle_{D,\,N}^{F}
&\stackrel{N}{=}&
\langle w^{\,\mu_{1}\,\ldots\,\mu_{m}}\rangle_{D,\,N}^{F}\;
\langle w^{\,\nu_{1}\,\ldots\,\nu_{n}}\rangle_{D,\,N}^{F}
\nonumber\\[1mm]
\hspace*{-3mm}&&
\, \cdots\,
\langle w^{\,\omega_{1}\,\ldots\,\omega_{z}}\rangle_{D,\,N}^{F}\,,
\eeqa
where the equality holds to leading order in $N$.
According to Witten~\cite{Witten1979}, the factorization
(\ref{eq:IIB-matrix-model-w-product-vev-factorized}) implies that
the path integrals (\ref{eq:IIB-matrix-model-w-product-vev}) are
saturated by a single configuration,
which has been called the ``master field''
and whose matrices will be denoted by $\widehat{A}^{\,\mu}$.
To leading order in $N$, the expectation values
\eqref{eq:IIB-matrix-model-w-product-vev} are then given by
the bosonic master-field matrices $\widehat{A}^{\,\mu}$
in the following way:%
\bsubeqs \label{eq:IIB-matrix-model-w-product-vev-from-master-field}
\beqa
\hspace*{-14.00mm}
&&\langle
w^{\mu_{1}\,\ldots\,\mu_{m}}\:w^{\nu_{1}\,\ldots\,\nu_{n}}\, \cdots\,
w^{\omega_{1}\,\ldots\,\omega_{z}} \rangle_{D,\,N}^{F}
\stackrel{N}{=}
\widehat{w}^{\,\mu_{1}\,\ldots\,\mu_{m}}\:
\widehat{w}^{\,\nu_{1}\,\ldots\,\nu_{n}}\, \cdots\,
\widehat{w}^{\,\omega_{1}\,\ldots\,\omega_{z}},
\\[2mm]
\hspace*{-14.00mm}&&
\widehat{w}^{\,\mu_{1}\,\ldots\, \mu_{m}}
\equiv
\frac{1}{N}\;
\text{Tr}\,\big( \widehat{A}^{\,\mu_{1}} \cdots\, \widehat{A}^{\,\mu_{m}}\big),
\eeqa
\esubeqs
where the master-field matrices $\widehat{A}^{\,\mu}$
have an implicit dependence on the model parameters
$D$, $N$, and $F$.
See Refs.~\cite{Klinkhamer2020-master,Klinkhamer2021-APPB-review}
for further discussion and references.


\subsection{Bosonic master-field equation}
\label{subsec:Bosonic-master-field-equation}

Introducing $N$ random constants $\widehat{p}_{k}$
and the $N \times N$ diagonal matrix
\beq
\label{eq:def-D}
D_{(\,\widehat{p}\,)}(\tau)  \equiv
\text{diag}
\left(
e^{\displaystyle{i\,\widehat{p}_{1}\,\tau}},\, \,\ldots\, ,\,
e^{\displaystyle{i\,\widehat{p}_{N}\,\tau}}
\right)\,,
\eeq
the bosonic master-field matrices take
the following ``quenched''
form~\cite{GreensiteHalpern1983,Klinkhamer2020-master}:
\bsubeqs\label{eq:Ahatrho-with-D-algebraic-equation-with-D}
\beqa\label{eq:Ahatrho-with-D}
\widehat{A}^{\;\rho}
&=&
D_{(\,\widehat{p}\,)}(\tau_\text{eq})
\cdot \widehat{a}^{\;\rho}
\cdot D_{(\,\widehat{p}\,)}^{-1}(\tau_\text{eq})\,,
\eeqa
for a sufficiently large value of $\tau_\text{eq}$
(see below for further explanations).
The $\tau$-independent matrix $\widehat{a}^{\;\rho}$
in \eqref{eq:Ahatrho-with-D} is determined by the algebraic
equation~\cite{Klinkhamer2020-master}
\beqa
\label{eq:algebraic-equation-with-D}
\frac{d}{d \tau}\,
\Big[D_{(\,\widehat{p}\,)}(\tau)
\cdot \widehat{a}^{\;\rho}
\cdot D_{(\,\widehat{p}\,)}^{-1}(\tau)\Big]_{\tau=0}
&=&
-\frac{\delta S_{\text{eff},\,D,\,N}^{F}\big[\,\widehat{a}\,\big]}
      {\delta\, \widehat{a}_{\rho}}\;
+\widehat{\eta}^{\;\rho}\,.
\eeqa
\esubeqs
All matrix indices have been suppressed in the three equations above
[the notation with a functional derivative on the right-hand side of
\eqref{eq:algebraic-equation-with-D} is purely symbolic]
and $S_\text{eff,\,D,\,N}^{F}[\,\widehat{a}\,]$
is given by \eqref{eq:Seff-D-N-F} or by
\eqref{eq:Seff-D3-N3} for the particular case considered.
The left-hand side of \eqref{eq:algebraic-equation-with-D},
with matrix indices $\{k,\,l\}$ added, reads
$i\,(\widehat{p}_{k}-\widehat{p}_{l})\,\widehat{a}^{\;\rho}_{\;kl}$
and the equation is manifestly algebraic (see
below for further comments on its basic structure).

The algebraic equation \eqref{eq:algebraic-equation-with-D}
has two types of constants:
the master momenta $\widehat{p}_{k}$ (uniform random numbers)
and the master noise  matrices $\widehat{\eta}^{\;\rho}_{\;kl}$
(Gaussian random numbers). Very briefly, the meaning
of these two types of random numbers is as follows.
The dimensionless time $\tau$ is the fictitious
Langevin time of stochastic quantization, with a
Gaussian noise term $\eta$ in the differential equation
(its basic structure is as follows:
$d A/d \tau= -\delta S_{\text{eff}}/\delta A +\eta$).
The $\tau$ evolution drives the system to equilibrium at
$\tau=\tau_\text{eq}$ and the resulting configuration
$A^{\rho}(\tau_\text{eq})$ corresponds to
the master field $\widehat{A}^{\;\rho}$.
For large $N$, the $\tau$-dependence of the bosonic variable
$A^{\rho}_{\,k\,l}(\tau)$
and the Langevin noise matrix $\eta^{\rho}_{\,k\,l}(\tau)$
is quenched by use of the uniform random momenta $\widehat{p}_{k}$.
Note that all master variables and master constants are denoted
by a caret. See Refs.~\cite{GreensiteHalpern1983,Klinkhamer2020-master}
for further discussion and references.

We now have three technical remarks on the
obtained master-field equations \eqref{eq:Ahatrho-with-D}
and \eqref{eq:algebraic-equation-with-D}.
First, the algebraic equation \eqref{eq:algebraic-equation-with-D}
for $F=0$ reproduces the simplified equation (3.1) of
Ref.~\cite{Klinkhamer2021-first-look},
up to an irrelevant minus sign of the double-commutator there.
Second, the derivative term
on the right-hand side of \eqref{eq:algebraic-equation-with-D}
for $F=1$ involves not only the derivative of the Pfaffian
(which has been studied in, e.g., Ref.~\cite{NishimuraVernizzi2000-JHEP})
but also the inverse of the Pfaffian.
Third, as the Pfaffian is a $K$-th order polynomial,
denoted symbolically by $P_{K}[A]$
with $K$ given by \eqref{eq:Pfaffian-order-polynomial-K},
the basic structure of the algebraic
equation \eqref{eq:algebraic-equation-with-D} is as follows:
\beqa
\label{eq:algebraic-equation-with-D-structure}
P_{1}^{(\,\widehat{p}\,)}\left[\widehat{a}\,\right]
&=&
P_{3}\left[\widehat{a}\,\right]
+F\,\frac{P_{K-1}\left[\widehat{a}\,\right]}{P_{K}\left[\widehat{a}\,\right]} 
+P_{0}^{(\,\widehat{\eta}\,)}\left[\widehat{a}\,\right]\,,
\eeqa
where only the on/off constant $F$ is shown explicitly
and where the suffixes on $P_{1}$ and $P_{0}$
indicate their respective dependence
on the master momenta $\widehat{p}_{k}$
and the master noise $\widehat{\eta}^{\;\rho}_{kl}$.
If we multiply \eqref{eq:algebraic-equation-with-D-structure} by
$P_{K}\left[\widehat{a}\,\right]$, we get a polynomial
equation of order $K+3$.

In order to obtain the component equations
[labelled by an index $I$ running over $1,\, \ldots \,,\, (N^2-1)$
and an index $\rho$ running over $1,\, \ldots \,,\, D$],
we matrix multiply \eqref{eq:algebraic-equation-with-D}
by $T_{I}$, take the trace, and multiply the result
by two.
There are then $D\,g=D\,\big(N^2-1\big)$ coupled algebraic equations
for an equal number of unknowns
$\{\widehat{a}_{1}^{\;1},\, \,\ldots\, ,\, \widehat{a}_{D}^{\;g}\}$.
It appears impossible to obtain a general solution of
these algebraic equations. We will
look for solutions of these coupled algebraic equations
with an explicit realization of the
random  constants $\widehat{p}_{k}$
and $\widehat{\eta}^{\;\rho}_{\;kl}$.
This is still a formidable problem for large values
of $N$. Only for very small values of $N$ are we, at this
moment, able to get an explicit result.

\section{Solutions for $D=3$ and $N=3$}
\label{sec:Solutions-D3-N3}

\subsection{Setup and method}
\label{subsec:Setup}

We will now obtain, for the particular case \eqref{eq:D3-N3},
several solutions of the algebraic equation
\eqref{eq:algebraic-equation-with-D}, which corresponds to
24 real algebraic equations for 24 real unknowns
$\{\widehat{a}_{1}^{\;1},\, \,\ldots\, ,\, \widehat{a}_{3}^{\;8}\}$.

For the constants entering these 24 algebraic equations,
we take pseudorandom rational numbers with a range $[-1/2,\,1/2]$
for the master momenta $\widehat{p}_{k}$
and pseudorandom rational numbers with a range $[-1,\,1]$
for the master noise coefficients $\widehat{\eta}_{\;\rho}^{\;I}$.
Specifically, we restrict to rational numbers of the
form $n/1000$, for $n \in \mathbb{Z}$, and take the
$\widehat{p}_{k}$ numbers from a uniform distribution
with a range $[-1/2,\,1/2]$ and
the $\widehat{\eta}_{\;\rho}^{\;I}$ numbers
from a truncated Gaussian distribution
(with spread $\sigma=2$ and cut-off value
$x_\text{trunc}=1$ in the notation of Sec.~III B of
Ref.~\cite{Klinkhamer2021-first-look};
ultimately we must take $x_\text{trunc}\gg \sigma$).

Explicitly, we use the following realization
(labeled $\alpha$) of the pseudorandom constants:
\bsubeqs\label{eq:pseudorandom-constants-D3-N3}
\beqa
\hspace*{-9.00mm}&& \widehat{p}_\text{\;$\alpha$-realization}
=
\left\{-\frac{27}{100}\,,\;\frac{257}{1000}\,,\;\frac{121}{1000}\right\}\,,
\eeqa
\beqa
\hspace*{-9.00mm}&&\widehat{\eta}^{\;1}_\text{\;$\alpha$-realization}
=
\nonumber\\[1mm]\hspace*{-9.00mm}&&
\left(
\renewcommand{\arraycolsep}{0.75pc} 
\renewcommand{\arraystretch}{2.5}  
  \begin{array}{ccc}
\fraclarge{547}{2000} + \fraclarge{13\,{\sqrt{3}}}{400} &
\fraclarge{36}{125} - \fraclarge{38\, i}{125} &
\fraclarge{103}{250} - \fraclarge{131\, i}{2000} \\
\fraclarge{36}{125} + \fraclarge{38\, i}{125} &
- \fraclarge{547}{2000}   + \fraclarge{13\, {\sqrt{3}}}{400} &
- \fraclarge{247}{2000}   + \fraclarge{41\, i}{250} \\
\fraclarge{103}{250} + \fraclarge{131\, i}{2000} &
- \fraclarge{247}{2000}   - \fraclarge{41\, i}{250} &
\fraclarge{-13\, {\sqrt{3}}}{200} \\
  \end{array}
\right)\,,
\eeqa
\beqa
\hspace*{-9.00mm}&&\widehat{\eta}^{\;2}_\text{\;$\alpha$-realization}
=
\nonumber\\[1mm]\hspace*{-9.00mm}&&
\left(
\renewcommand{\arraycolsep}{0.75pc} 
\renewcommand{\arraystretch}{2.5}  
  \begin{array}{ccc}
\fraclarge{319}{2000} - \fraclarge{467}{2000\, {\sqrt{3}}} &
-  \fraclarge{921}{2000}   + \fraclarge{43\, i}{400} &
\fraclarge{163}{1000} + \fraclarge{97\, i}{400} \\
-  \fraclarge{921}{2000}   - \fraclarge{43\, i}{400} &
-  \fraclarge{319}{2000}   - \fraclarge{467}{2000\, {\sqrt{3}}} &
\fraclarge{951}{2000} + \fraclarge{419\, i}{2000} \\
\fraclarge{163}{1000} - \fraclarge{97\, i}{400} &
\fraclarge{951}{2000} - \fraclarge{419\, i}{2000} & \fraclarge{467}{1000\, {\sqrt{3}}} \\
 \end{array}
\right)\,,
\eeqa
\beqa
\hspace*{-9.00mm}&&\widehat{\eta}^{\;3}_\text{\;$\alpha$-realization}
=
\nonumber\\[1mm]\hspace*{-9.00mm}&&
\left(
\renewcommand{\arraycolsep}{0.75pc} 
\renewcommand{\arraystretch}{2.5}  
  \begin{array}{ccc}
\fraclarge{28}{125} + \fraclarge{989}{2000\, {\sqrt{3}}} &
- \fraclarge{419}{1000}   + \fraclarge{27\, i}{1000} &
\fraclarge{169}{500} + \fraclarge{13\, i}{40} \\
-  \fraclarge{419}{1000}   - \fraclarge{27\, i}{1000} &
-  \fraclarge{28}{125}   + \fraclarge{989}{2000\, {\sqrt{3}}} &
\fraclarge{219}{500} - \fraclarge{241\, i}{1000} \\
\fraclarge{169}{500} - \fraclarge{13\, i}{40} &
\fraclarge{219}{500}+\fraclarge{241\, i}{1000} &
\fraclarge{-989}{1000\, {\sqrt{3}}} \\
 \end{array}
\right)\,.
\eeqa
\esubeqs
The tracelessness of the $\widehat{\eta}^{\;\rho}$ matrices
in \eqref{eq:pseudorandom-constants-D3-N3} is manifest.
Other realizations
(labeled $\beta,\,\gamma,\, \ldots$) have given similar results.

The 3 solution matrices
$\widehat{a}_{\alpha\text{-sol}}^{\;\rho}$
are determined by 24 coefficients
$\big(\widehat{a}_{\alpha\text{-sol}}^{\;\rho}\big)^{I}$.
Before we present these coefficients,
which are obtained from the 24 algebraic equations mentioned above,
let us briefly describe the method used.
Strictly speaking, it does not matter how
the 24 real numbers $\big(\widehat{a}_{\alpha\text{-sol}}^{\;\rho}\big)^{I}$
are obtained, as long as they solve the 24 algebraic equations.
Here, we obtain these 24 real numbers
with the numerical minimization routine \texttt{FindMinimum}
from \textsc{Mathematica} 12.1
(cf. Ref.~\cite{Wolfram1991}).
The minimization operates on a penalty function,
which consists of a
sum of 24 squares, each square containing
one of the real components of the algebraic equation
without further overall numerical factor.
(This penalty function has a size of approximately $29$ MB,
as further simplifications are hard to obtain.)
For all calculations of the present paper,
we use a 36-digit working precision.
The accuracy of the obtained 24 real numbers
$\big(\widehat{a}_{\alpha\text{-sol}}^{\;\rho}\big)^{I}$
can, in principle, be increased arbitrarily.
Given the exact (pseudorandom) rational
constants
$\widehat{p}_{k}$  and
$\widehat{\eta}_{\;\rho}^{\;I}$
from \eqref{eq:pseudorandom-constants-D3-N3}, the
obtained matrices may, therefore, be called ``quasi-exact.''


\subsection{Solution without dynamic fermions ($F=0$)}
\label{subsec:Solution-without-dynamic-fermions}


We, first, get a solution of the $D=N=3$ algebraic master-field
equation \eqref{eq:algebraic-equation-with-D},
where dynamic-fermion effects have been excluded
by setting $F=0$ in the effective action \eqref{eq:Seff-D3-N3}.
Then, the resulting 24 coupled algebraic equations,
with constants \eqref{eq:pseudorandom-constants-D3-N3},
have the following solution (for display and readability reasons,
we split each matrix into the sum of a matrix with real entries
and a matrix with imaginary entries):
\bsubeqs\label{eq:ahat1numsol-ahat2numsol-ahat3numsol-N3-F0}
\beqa
\label{eq:ahat1numsol-N3-F0}
\hspace*{-9.00mm}
\widehat{a}^{\;1}_\text{\;$\alpha$-sol}\,
\Big|^{(F=0)}
&=&
\left(
\renewcommand{\arraycolsep}{0.5pc} 
\renewcommand{\arraystretch}{1.5}  
  \begin{array}{ccc}
0.186159 & 0.073147   & 0.562726   \\
0.073147   & - 0.281384 & - 0.393265  \\
0.562726   & - 0.393265  & 0.0952246
  \end{array}
\right)
\nonumber\\[1mm]
\hspace*{-9.00mm}&&
+
\left(
\renewcommand{\arraycolsep}{0.5pc} 
\renewcommand{\arraystretch}{1.5}  
  \begin{array}{ccc}
0 &  0.401829\, i  & - 0.217741\, i  \\
 - 0.401829\, i  &  0 &  - 0.100372\, i  \\
0.217741\, i  &  0.100372\, i  & 0
  \end{array}
\right),
\eeqa
\beqa
\label{eq:ahat2numsol-N3-F0}
\hspace*{-9.00mm}
\widehat{a}^{\;2}_\text{\;$\alpha$-sol}\,\Big|^{(F=0)}
&=&
\left(
\renewcommand{\arraycolsep}{0.5pc} 
\renewcommand{\arraystretch}{1.5}  
  \begin{array}{ccc}
-0.194103 & 0.008581  & - 0.580188   \\
0.008581   & 0.131909 & 0.546198   \\
-0.580188   & 0.546198   & 0.0621938
  \end{array}
\right)
\nonumber\\
\hspace*{-9.00mm}&&
+
\left(
\renewcommand{\arraycolsep}{0.5pc} 
\renewcommand{\arraystretch}{1.5}  
  \begin{array}{ccc}
0 &  - 0.360539\, i  &  0.482773\, i  \\
 0.360539\, i  & 0 &  0.389392\, i  \\
 - 0.482773\, i  &  - 0.389392\, i  & 0
  \end{array}
\right)\,,
\eeqa
\beqa
\label{eq:ahat3numsol-N3-F0}
\hspace*{-9.00mm}
\widehat{a}^{\;3}_\text{\;$\alpha$-sol}\,\Big|^{(F=0)}
&=&
\left(
\renewcommand{\arraycolsep}{0.5pc} 
\renewcommand{\arraystretch}{1.5}  
  \begin{array}{ccc}
0.130861 & - 0.201837   & - 0.1076245  \\
- 0.201837   & 0.304673 & 0.277152  \\
- 0.1076245   & 0.277152   & - 0.435534
  \end{array}
\right)
\nonumber\\
\hspace*{-9.00mm}&&
+
\left(
\renewcommand{\arraycolsep}{0.5pc} 
\renewcommand{\arraystretch}{1.5}  
  \begin{array}{ccc}
0 &  - 0.581365\, i  & - 0.0760187\, i \\
 0.581365\, i  & 0 &  - 0.281138\, i  \\
 0.0760187\, i  &  0.281138\, i  &  0
  \end{array}
\right)\,,
\eeqa
\esubeqs
where up to 6 significant digits are shown
(a 36-digit working precision is used).
The apparent violation of tracelessness is solely due
to roundoff errors and is absent in the true solution, which
is given by an expansion in terms of traceless generators,
as in \eqref{eq:Amu-coeff-Psialpha-coeff}.

A cursory inspection of the
matrices \eqref{eq:ahat1numsol-ahat2numsol-ahat3numsol-N3-F0}
shows that the far-off-diagonal
entries $[1,\,3]$ and $[3,\,1]$ are not all really
small. Following the discussion of our previous
paper~\cite{Klinkhamer2021-first-look}, we will
consider the absolute values of the entries in the
$\rho=1$ matrix \eqref{eq:ahat1numsol-N3-F0}, calculate
the average band-diagonal value from 2+3+2 entries,
the average off-band-diagonal value from 1+1 entries,
and the ratio $R_{1}$ of the average band-diagonal value
over the average off-band-diagonal value.
For the $\rho=2$ and $\rho=3$ matrices,
we follow the same procedure and get the ratios
$R_{2}$ and $R_{3}$. In order to avoid any confusion,
we give the general definition of this ratio $R$ for a symmetric
$3\times 3$ matrix $M$ with nonnegative entries $m[i,\,j]$:  
\beq
\label{eq:def-R-M}
R_{M} \equiv
\frac{1}{7}\,\left( \sum_{i=1}^{3} m[i,\,i] +2\,\sum_{j=1}^{2} m[j,\,j+1]\right)\;
\frac{1}{m[1,\,3]}\,,
\eeq
where the symmetry of $M$ has been used to simplify the expression.

From the matrices \eqref{eq:ahat1numsol-ahat2numsol-ahat3numsol-N3-F0},
we then get the following ratio values:
\beqa\label{eq:amunumsol-N3-F0-ratios}
\hspace*{-0mm}
\Big\{
R_{1},\,R_{2},\,R_{3}
\Big\}_{\text{Abs}\left[\widehat{a}^{\;\rho}_\text{\;$\alpha$-sol}\right]}^{(F=0)}
&\!=\!&
\left\{ 0.519,\,0.464,\,3.13 \right\}\,,
\eeqa
where two ratios lie below unity and one above.
The somewhat large value of the last ratio
in \eqref{eq:amunumsol-N3-F0-ratios} is
due to the rather small $[1,\,3]$ and $[3,\,1]$ entries in
the matrix \eqref{eq:ahat3numsol-N3-F0}.

Next, we diagonalize one of the matrices, while ordering the eigenvalues,
and look at the other two matrices
to see if they have a band-diagonal structure
(even for the very small value of $N$ we are considering).
If we diagonalize and order $\widehat{a}^{\;1}_\text{\;$\alpha$-sol}$
(the new matrices are denoted by a prime), we get
\bsubeqs\label{eq:amuprimenumsol-N3-F0}
\beqa
\label{eq:a1primenumsol-N3-F0}
\hspace*{-9.00mm}
\widehat{a}^{\;\prime\,1\;(F=0)}_\text{\;$\alpha$-sol}
&=&
S_{1} \cdot \widehat{a}^{\;1\;(F=0)}_\text{\;$\alpha$-sol}
  \cdot S_{1}^{-1}
\nonumber\\[1mm]
\hspace*{-9.00mm}
&=&
\text{diag}\big(\! -0.760473,\,-0.188386,\,0.948859\big)\,,
\\[2mm]
\label{eq:a2primenumsol-N3-F0}
\hspace*{-9.00mm}
\widehat{a}^{\;\prime\,2\;(F=0)}_\text{\;$\alpha$-sol}
&=& S_{1} \cdot \widehat{a}^{\;2\;(F=0)}_\text{\;$\alpha$-sol}
\cdot S_{1}^{-1}
\nonumber\\[1mm]
\hspace*{-9.00mm}&=&
\left(
\renewcommand{\arraycolsep}{0.5pc} 
\renewcommand{\arraystretch}{1.5}  
  \begin{array}{ccc}
0.666791 & - 0.102279   & - 0.0133662  \\
- 0.102279   & 0.516194 & 0.066637   \\
- 0.0133662   & 0.066637  & - 1.18298 \\
   \end{array}
\right)
\nonumber\\[1mm]
\hspace*{-9.00mm}&&
+
\left(
\renewcommand{\arraycolsep}{0.5pc} 
\renewcommand{\arraystretch}{1.5}  
  \begin{array}{ccc}
0  &  - 0.255347\, i  &  0.0232436\, i \\
 0.255347\, i    &  0 &  0.207603\, i  \\
 -0.0232436\, i  &  -0.207603\, i  & 0 \\
   \end{array}
\right),
\eeqa
\beqa
\label{eq:a3primenumsol-N3-F0}
\hspace*{-9.00mm}
\widehat{a}^{\;\prime\,3\;(F=0)}_\text{\;$\alpha$-sol}
&=& S_{1} \cdot \widehat{a}^{\;3\;(F=0)}_\text{\;$\alpha$-sol} \cdot S_{1}^{-1}
\nonumber\\[1mm]
\hspace*{-9.00mm}&=&
\left(
\renewcommand{\arraycolsep}{0.25pc} 
\renewcommand{\arraystretch}{1.5}  
  \begin{array}{ccc}
0.727690 & 0.273460   & - 0.0776651   \\
0.273460   & - 0.347068 & 0.136913  \\
- 0.0776651   & 0.136913   & - 0.380622 \\
   \end{array}
\right)
\nonumber\\
\hspace*{-9.00mm}&&
+
\left(
\renewcommand{\arraycolsep}{0.25pc} 
\renewcommand{\arraystretch}{1.5}  
  \begin{array}{ccc}
0 &  0.440485\, i  &  0.0239532\, i  \\
 - 0.440485\, i  &  0 &  0.100836\, i  \\
 - 0.0239532\, i  & - 0.100836\, i  &  0 \\
   \end{array}
\right).
\eeqa
\esubeqs
The last two matrices in \eqref{eq:amuprimenumsol-N3-F0}
have a rather clear band-diagonal structure.
This can, again, be quantified by the
ratios $R_{\rho}$ of the average band-diagonal value
over the average  off-band-diagonal value,
\beqa\label{eq:amuprimenumsol-N3-F0-ratios}
\hspace*{-0mm}
\Big\{
R_{1},\,R_{2},\,R_{3}
\Big\}_{\text{Abs}\left[\widehat{a}^{\;\prime\,\rho}_\text{\;$\alpha$-sol}\right]}^{(F=0)}
&\!=\!&
\left\{ \infty,\,  17.9,\, 4.98 \right\}\,,
\eeqa
where the first ratio is simply infinite because the $[1,\,3]$
entry of the matrix \eqref{eq:a1primenumsol-N3-F0} vanishes.
All three ratios from \eqref{eq:amuprimenumsol-N3-F0-ratios}
are larger than 1, just
as seen in our previous results~\cite{Klinkhamer2021-first-look}.


\subsection{Solutions with dynamic fermions ($F=1$)}
\label{subsec:Solutions-with-dynamic-fermions}

We, next, get solutions of the $D=N=3$ algebraic master-field
equation \eqref{eq:algebraic-equation-with-D},
where dynamic-fermion effects have been included
by setting $F=1$ in the effective action \eqref{eq:Seff-D3-N3}.
The resulting 24 coupled algebraic equations,
with constants \eqref{eq:pseudorandom-constants-D3-N3},
have then the following two solutions.

\subsubsection{First solution}
\label{subsubsec:First-solution}


Starting the minimization procedure from
the configuration \eqref{eq:ahat1numsol-ahat2numsol-ahat3numsol-N3-F0}
obtained in Sec.~\ref{subsec:Solution-without-dynamic-fermions},
we find
\bsubeqs\label{eq:ahat1numsol-ahat2numsol-ahat3numsol-N3-F1}
\beqa
\label{eq:ahat1numsol-N3-F1}
\hspace*{-9.00mm}\widehat{a}^{\;1}_\text{\;$\alpha$-sol}\,\Big|^{(F=1)}
&=&
\left(
\renewcommand{\arraycolsep}{0.5pc} 
\renewcommand{\arraystretch}{1.5}  
  \begin{array}{ccc}
0.481805 & 0.325813   & 0.384395  \\
0.325813   & - 0.682181 & - 0.183233  \\
0.384395   & - 0.183233   & 0.200375 \\
  \end{array}
\right)
\nonumber\\[1mm]\hspace*{-9.00mm}&&
+
\left(
\renewcommand{\arraycolsep}{0.5pc} 
\renewcommand{\arraystretch}{1.5}  
  \begin{array}{ccc}
0 &  0.576164\, i  & - 0.050542\, i  \\
 - 0.576164\, i  &  0 & 0.593735\, i \\
 0.050542\, i  &  - 0.593735\, i  & 0 \\
  \end{array}
\right),
\eeqa
\beqa
\label{eq:ahat2numsol-N3-F1}
\hspace*{-9.00mm}\widehat{a}^{\;2}_\text{\;$\alpha$-sol}\,\Big|^{(F=1)}
&=&
\left(
\renewcommand{\arraycolsep}{0.5pc} 
\renewcommand{\arraystretch}{1.5}  
  \begin{array}{ccc}
-0.222436 & - 0.0458796   & 0.048921 \\
- 0.0458796   & 0.000171969 & 0.287881   \\
0.048921   & 0.287881   & 0.222264 \\
  \end{array}
\right)
\nonumber\\[1mm]\hspace*{-9.00mm}&&
+
\left(
\renewcommand{\arraycolsep}{0.5pc} 
\renewcommand{\arraystretch}{1.5}  
  \begin{array}{ccc}
0 &  - 0.1004043\, i  &  0.503341\, i \\
 0.1004043\, i  & 0 & 0.093830\, i  \\
 - 0.503341\, i  &  -0.093830\, i  & 0 \\
  \end{array}
\right)\,,
\eeqa
\beqa
\label{eq:ahat3numsol-N3-F1}
\hspace*{-9.00mm}\widehat{a}^{\;3}_\text{\;$\alpha$-sol}\,\Big|^{(F=1)}
&=&
\left(
\renewcommand{\arraycolsep}{0.5pc} 
\renewcommand{\arraystretch}{1.5}  
  \begin{array}{ccc}
0.0389571 & - 0.304651  & - 0.194354   \\
- 0.304651   & 0.858818 & 0.777316   \\
- 0.194354   & 0.777316   & -0.897775 \\
  \end{array}
\right)\
\nonumber\\[1mm]\hspace*{-9.00mm}&&
+
\left(
\renewcommand{\arraycolsep}{0.5pc} 
\renewcommand{\arraystretch}{1.5}  
  \begin{array}{ccc}
0 &  - 0.920567\, i  & 0.239810\, i  \\
 0.920567\, i  & 0 &  - 0.977785\, i  \\
 - 0.239810\, i  &  0.977785\, i  & 0 \\
  \end{array}
\right)\,,
\eeqa
\esubeqs
where up to 6 significant digits are shown
(as mentioned before, the apparent violation of tracelessness is due
to roundoff errors).
The most important result of the present paper is
the fact that we were able to find a
nontrivial solution \eqref{eq:ahat1numsol-ahat2numsol-ahat3numsol-N3-F1}
for the case of dynamic fermions (another solution will
be given in Sec.~\ref{subsubsec:Second-solution}).

There is perhaps a slight resemblance between the
$F=0$ matrices \eqref{eq:ahat1numsol-ahat2numsol-ahat3numsol-N3-F0}
and the
$F=1$ matrices \eqref{eq:ahat1numsol-ahat2numsol-ahat3numsol-N3-F1},
if, for example, the signs of the diagonal elements are considered.
Anyway, it is clear that
the far-off-diagonal entries $[1,\,3]$ and $[3,\,1]$
of the matrices \eqref{eq:ahat1numsol-ahat2numsol-ahat3numsol-N3-F1}
are not really small.
Again, we will consider the absolute values of the entries in
the matrices \eqref{eq:ahat1numsol-ahat2numsol-ahat3numsol-N3-F1}
and calculate, for each matrix, the
ratio $R$ of the average band-diagonal value
over the average off-band-diagonal value,
according to the general definition \eqref{eq:def-R-M}.
We then get the following ratio values:
\beqa\label{eq:amunumsol-N3-F1-ratios}
\hspace*{-0mm}
\Big\{
R_{1},\,R_{2},\,R_{3}
\Big\}_{\text{Abs}\left[\widehat{a}^{\;\rho}_\text{\;$\alpha$-sol}\right]}^{(F=1)}
&\!=\!&
\left\{ 1.45,\, 0.359,\, 2.88 \right\}\,,
\eeqa
where two ratios lie above unity and one below.

Next, we diagonalize and order $\widehat{a}^{\;1\;(F=1)}_\text{\;$\alpha$-sol}$
(the new matrices are denoted by a prime) and get
\bsubeqs\label{eq:amuprimenumsol-N3-F1}
\beqa
\hspace*{-9.00mm}\widehat{a}^{\;\prime\,1\;(F=1)}_\text{\;$\alpha$-sol}
&=& \widetilde{S}_{1} \cdot \widehat{a}^{\;1\;(F=1)}_\text{\;$\alpha$-sol}
\cdot \widetilde{S}_{1}^{-1}
\nonumber\\[1mm]
\hspace*{-9.00mm}&=&
\text{diag}\big(\!  -1.32720,\, 0.514388,\, 0.812811 \big)\,,
\eeqa
\beqa
\label{eq:a2primenumsol-N3-F1}
\hspace*{-9.00mm}\widehat{a}^{\;\prime\,2\;(F=1)}_\text{\;$\alpha$-sol}
&=& \widetilde{S}_{1} \cdot \widehat{a}^{\;2\;(F=1)}_\text{\;$\alpha$-sol}
\cdot \widetilde{S}_{1}^{-1}
\nonumber\\[1mm]
\hspace*{-9.00mm}&=&
\left(
\renewcommand{\arraycolsep}{0.5pc} 
\renewcommand{\arraystretch}{1.5}  
  \begin{array}{ccc}
0.0919594 & 0.1408718   & - 0.177064   \\
0.1408718   & 0.0633399 & 0.179233   \\
- 0.177064  & 0.179233   & -0.155299 \\
   \end{array}
\right)
\nonumber\\[1mm]
\hspace*{-9.00mm}&&
+
\left(
\renewcommand{\arraycolsep}{0.5pc} 
\renewcommand{\arraystretch}{1.5}  
  \begin{array}{ccc}
0 &  - 0.0118767\, i  &  - 0.134760\, i  \\
 0.0118767\, i  & 0 &  - 0.537712\, i  \\
 0.134760\, i  & 0.537712\, i  & 0 \\
   \end{array}
\right),
\eeqa
\beqa
\label{eq:a3primenumsol-N3-F1}
\hspace*{-9.00mm}\widehat{a}^{\;\prime\,3\;(F=1)}_\text{\;$\alpha$-sol}
&=& \widetilde{S}_{1} \cdot \widehat{a}^{\;3\;(F=1)}_\text{\;$\alpha$-sol}
\cdot \widetilde{S}_{1}^{-1}
\nonumber\\[1mm]
\hspace*{-9.00mm}&=&
\left(
\renewcommand{\arraycolsep}{0.5pc} 
\renewcommand{\arraystretch}{1.5}  
  \begin{array}{ccc}
1.91375 & 0.245877   & 0.022834   \\
0.245877  & - 1.57339 & 0.102361  \\
0.022834   & 0.102361   & - 0.340362 \\
   \end{array}
\right)
\nonumber\\[1mm]
\hspace*{-9.00mm}&&
+
\left(
\renewcommand{\arraycolsep}{0.5pc} 
\renewcommand{\arraystretch}{1.5}  
  \begin{array}{ccc}
0 &  0.219132\, i  &  0.184308\, i  \\
 - 0.219132\, i  & 0 &  - 0.296644\, i  \\
 - 0.184308\, i  &  0.296644\, i  &  0 \\
   \end{array}
\right).
\eeqa
\esubeqs
The last two matrices in \eqref{eq:amuprimenumsol-N3-F1}
have a mild band-diagonal structure, quantified by the following
ratios $R_{\rho}$ of the average band-diagonal value
over the average  off-band-diagonal value:
\beqa\label{eq:amuprimenumsol-N3-F1-ratios}
\hspace*{-0mm}
\Big\{
R_{1},\,R_{2},\,R_{3}
\Big\}_{\text{Abs}\left[\widehat{a}^{\;\prime\,\rho}_\text{\;$\alpha$-sol}\right]}^{(F=1)}
&\!=\!&
\left\{ \infty,\,   1.11,\, 3.93  \right\}\,.
\eeqa
All three values in \eqref{eq:amuprimenumsol-N3-F1-ratios}
lie above unity, but the second not by much.


\subsubsection{Second solution}
\label{subsubsec:Second-solution}

With a different start configuration, we obtain another dynamic-fermion
solution (denoted by an underline):  
\bsubeqs\label{eq:second-ahat1numsol-ahat2numsol-ahat3numsol-N3-F1}
\beqa
\label{eq:second-ahat1numsol-N3-F1}
\hspace*{-11.00mm}
\underline{\widehat{a}}^{\;1}_\text{\;$\alpha$-sol}\,\Big|^{(F=1)}
&=&
\left(
\renewcommand{\arraycolsep}{0.5pc} 
\renewcommand{\arraystretch}{1.5}  
  \begin{array}{ccc}
0.125763 & 0.412706   & 0.233008   \\
0.412706   & - 0.817158 & 0.126363   \\
0.233008   & 0.126363   & 0.691395 \\
  \end{array}
\right)
\nonumber\\
\hspace*{-11.00mm}&&
+
\left(
\renewcommand{\arraycolsep}{0.5pc} 
\renewcommand{\arraystretch}{1.5}  
  \begin{array}{ccc}
0 &  - 0.424807\, i  &  - 0.106924\, i  \\
 0.424807\, i  &  0 &  - 0.774279\, i  \\
 0.106924\, i  &  0.774279\, i  & 0 \\
  \end{array}
\right),
\eeqa
\beqa
\label{eq:second-ahat2numsol-N3-F1}
\hspace*{-11.00mm}\underline{\widehat{a}}^{\;2}_\text{\;$\alpha$-sol}\,\Big|^{(F=1)}
&=&
\left(
\renewcommand{\arraycolsep}{0.5pc} 
\renewcommand{\arraystretch}{1.5}  
  \begin{array}{ccc}
-0.0889768 & - 0.256429   & 0.164899  \\
- 0.256429  & - 0.299159 & - 0.078519   \\
0.164899   & - 0.078519   & 0.388136 \\
  \end{array}
\right)
\nonumber\\
\hspace*{-11.00mm}&&
+
\left(
\renewcommand{\arraycolsep}{0.5pc} 
\renewcommand{\arraystretch}{1.5}  
  \begin{array}{ccc}
0 &  - 0.199471\, i  &   0.170884\, i  \\
 0.199471\, i  &  0 &  - 0.153213\, i  \\
 - 0.170884\, i  &  0.153213\, i  & 0 \\
  \end{array}
\right)\,,
\eeqa
\beqa
\label{eq:second-ahat3numsol-N3-F1}
\hspace*{-11.00mm}\underline{\widehat{a}}^{\;3}_\text{\;$\alpha$-sol}\,\Big|^{(F=1)}
&=&
\left(
\renewcommand{\arraycolsep}{0.5pc} 
\renewcommand{\arraystretch}{1.5}  
  \begin{array}{ccc}
0.823634 & - 0.164815   & - 0.353134   \\
- 0.164815   & -0.365779 & - 0.158931   \\
- 0.353134   & - 0.158931   & - 0.457855 \\
  \end{array}
\right)
\nonumber\\
\hspace*{-11.00mm}&&
+
\left(
\renewcommand{\arraycolsep}{0.5pc} 
\renewcommand{\arraystretch}{1.5}  
  \begin{array}{ccc}
0 &  - 0.354320\, i  &  0.320626\, i  \\
 0.354320\, i  & 0 &  - 0.293612\, i  \\
 - 0.320626\, i  &  0.293612\, i  &  0 \\
  \end{array}
\right)\,,
\eeqa
\esubeqs
where up to 6 significant digits are shown
(the apparent violation of tracelessness is due to roundoff errors).

Inspection of the
matrices \eqref{eq:second-ahat1numsol-ahat2numsol-ahat3numsol-N3-F1}
shows that the far-off-diagonal
entries $[1,\,3]$ and $[3,\,1]$ are not really small.
Considering the absolute values of the entries in
the matrices \eqref{eq:second-ahat1numsol-ahat2numsol-ahat3numsol-N3-F1}
and calculating, for each matrix, the
ratio $R$ of the average band-diagonal value
over the average off-band-diagonal value,
according to the general definition \eqref{eq:def-R-M},
we get the following ratio values:
\beqa\label{eq:second-amunumsol-N3-F0-ratios}
\hspace*{-0mm}
\Big\{
R_{1},\,R_{2},\,R_{3}
\Big\}_{\text{Abs}\left[\underline{\widehat{a}}^{\,\rho}_\text{\;$\alpha$-sol}\right]}^{(F=1)}
&\!=\!&
\left\{ 2.45,\, 1.06,\, 0.927 \right\}\,,
\eeqa
where all three ratios are of order unity.

Next, we diagonalize and order $\underline{\widehat{a}}^{\;1}_\text{\;$\alpha$-sol}$
(the new matrices are denoted by a further prime) and get
\bsubeqs\label{eq:second-amuprimenumsol-N3-F1}
\beqa
\hspace*{-9.00mm}\underline{\widehat{a}}^{\;\prime\,1\;(F=1)}_\text{\;$\alpha$-sol}
&=&
 \underline{S}_{\;1} \cdot \underline{\widehat{a}}^{\,1\;(F=1)}_\text{\;$\alpha$-sol}
 \cdot \underline{S}_{\;1}^{-1}
\nonumber\\[1mm]
\hspace*{-9.00mm}&=&
\text{diag}\big(\!  -1.37609,\,  0.249211,\, 1.12688 \big)\,,
\eeqa
\beqa
\label{eq:second-a2primenumsol-N3-F1}
\hspace*{-9.00mm}
\underline{\widehat{a}}^{\;\prime\,2\;(F=1)}_\text{\;$\alpha$-sol}
&=&
\underline{S}_{\;1}
\cdot \underline{\widehat{a}}^{\,2\;(F=1)}_\text{\;$\alpha$-sol}
\cdot \underline{S}_{\;1}^{-1}
\nonumber\\[1mm]
\hspace*{-9.00mm}&=&
\left(
\renewcommand{\arraycolsep}{0.5pc} 
\renewcommand{\arraystretch}{1.5}  
  \begin{array}{ccc}
-0.346970 & - 0.204388   & 0.0067429  \\
- 0.204388   & 0.172384 & 0.049531   \\
0.0067429  & 0.049531   & 0.174586 \\
   \end{array}
\right)
\nonumber\\[1mm]
\hspace*{-9.00mm}&&
+
\left(
\renewcommand{\arraycolsep}{0.5pc} 
\renewcommand{\arraystretch}{1.5}  
  \begin{array}{ccc}
0 &  0.161013\, i  &  0.1331020\, i \\
 - 0.161013\, i  & 0 &  0.370677\, i  \\
 - 0.1331020\, i  &  - 0.370677\, i  & 0 \\
   \end{array}
\right),
\eeqa
\beqa
\label{eq:second-a3primenumsol-N3-F1}
\hspace*{-9.00mm}\underline{\widehat{a}}^{\;\prime\,3\;(F=1)}_\text{\;$\alpha$-sol}
&=& \underline{S}_{\;1}
\cdot \underline{\widehat{a}}^{\,3\;(F=1)}_\text{\;$\alpha$-sol}
\cdot \underline{S}_{\;1}^{-1}
\nonumber\\[1mm]
\hspace*{-9.00mm}&=&
\left(
\renewcommand{\arraycolsep}{0.5pc} 
\renewcommand{\arraystretch}{1.5}  
  \begin{array}{ccc}
-0.491709 & - 0.366405   & - 0.0482257   \\
- 0.366405   & 0.909724 & - 0.428317  \\
- 0.0482257   & - 0.428317   & - 0.418015 \\
   \end{array}
\right)
\nonumber\\[1mm]
\hspace*{-9.00mm}&&
+
\left(
\renewcommand{\arraycolsep}{0.5pc} 
\renewcommand{\arraystretch}{1.5}  
  \begin{array}{ccc}
0 &  - 0.207799\, i  &  - 0.0663827\, i  \\
 0.207799\, i  & 0 &  - 0.113580\, i \\
 0.0663827\, i  &  0.113580\, i  &  0 \\
   \end{array}
\right).
\eeqa
\esubeqs
The last two matrices in \eqref{eq:second-amuprimenumsol-N3-F1}
have a mild band-diagonal structure.
This can, again, be quantified by the
ratios $R_{\rho}$ of the average band-diagonal value
over the average off-band-diagonal value:  
\beqa\label{eq:second-amuprimenumsol-N3-F1-ratios}
\hspace*{-0mm}
\Big\{
R_{1},\,R_{2},\,R_{3}
\Big\}_{\text{Abs}\left[\underline{\widehat{a}}^{\;\prime\,\rho}_\text{\;$\alpha$-sol}\right]}^{(F=1)}
&\!=\!&
\left\{ \infty,\, 2.10,\,  6.18 \right\}\,.
\eeqa
All three values in \eqref{eq:second-amuprimenumsol-N3-F1-ratios}
lie above unity, the last two values being somewhat larger than those  
of the first $F=1$ solution in \eqref{eq:amuprimenumsol-N3-F1-ratios}.


\section{Discussion}
\label{sec:Discussion}

In the present article, we have obtained, for the first time,
solutions of the full bosonic master-field equation
from the supersymmetric matrix model \eqref{eq:Z-D-N-F-all-defs}
with dimensionality \mbox{$D=3$,} matrix size $N=3$,
and fermion-inclusion parameter $F=1$.
These particular values of $D$ and $N$ are, of course,
far below the values \eqref{eq:D-N-F-for-IIB-matrix-model}
needed for the IIB matrix model~\cite{IKKT-1997,Aoki-etal-review-1999}.
Still, it is an important point of principle
to have established the existence of solutions
with full fermion dynamics,
even for small values of $D$ and $N$.
Let us turn the argument around: assume that there were
no such solutions for $D=3$ and $N=3$, then it would be
hard to believe that there could be solutions for $D=10$
and $N\gg 1$, as needed for the IIB matrix model.
For completeness, we mention that the
bosonic master-field equation \eqref{eq:algebraic-equation-with-D}
is purely algebraic, with a basic structure clarified
in \eqref{eq:algebraic-equation-with-D-structure}
for $K$ as given by \eqref{eq:Pfaffian-order-polynomial-K} .

The matrices of the obtained bosonic $D=N=3$ solutions feel
the induction effects of the dynamic fermions ($F=1$).
The comparison between the solution without dynamic fermions
(Sec.~\ref{subsec:Solution-without-dynamic-fermions})
and the corresponding solution with dynamic
fermions (Sec.~\ref{subsubsec:First-solution})
suggests that the fermions reduce
somewhat the strength of the diagonal/band-diagonal structure
residing in the obtained matrices [specifically, compare the
ratios in \eqref{eq:amuprimenumsol-N3-F0-ratios}
with those in \eqref{eq:amuprimenumsol-N3-F1-ratios}].
But there is, at least, one other dynamic-fermion
solution (Sec.~\ref{subsubsec:Second-solution}),
which has a somewhat
more pronounced diagonal/band-diagonal structure,
as shown by the ratios
in \eqref{eq:second-amuprimenumsol-N3-F1-ratios}.

At this moment, we would like to proceed to larger values of
($D,\,N$), in order to ultimately reach the parameter values
$D=10$ and $N \gg 1$ of the
genuine IIB matrix model~\cite{IKKT-1997,Aoki-etal-review-1999}.
But this will be difficult.
Already the matrix model with modest values $(D,\,N)=(4,\,4)$
has a Pfaffian given as the determinant of a
$30 \times 30$ complex matrix~\cite{KrauthNicolaiStaudacher1998}.
And the matrix model with $(D,\,N)=(10,\,4)$
has a Pfaffian of a  $240 \times 240$ skew-symmetric matrix.
In both cases, it will be hard to evaluate the Pfaffian symbolically.

Instead of searching for a direct algebraic   
solution of the full bosonic master-field equation,
an indirect numerical approach may be called for or even a reliable
approximation method, if at all available.

\section*{\hspace*{-4.5mm}Acknowledgments}

The referee is thanked for helpful remarks on Eqs.~\eqref{eq:mathcalN}
and \eqref{eq:Pfaffian-order-polynomial-K}.


\end{document}